# A Manifesto for a Pro-Actively Responsible AI in Education

**Kaśka Porayska-Pomsta**

Kaśka Porayska-Pomsta k.porayska-pomsta@ucl.ac.uk

WC1E 6BT London, UK

Faculty of Education and Society, UCL Knowledge Lab, University College London, Gower St.,

## Introduction

The field of AIED, as defined by the work conducted under the auspices of the International Society of Artificial Intelligence in Education, has been built on big and well-intentioned ambitions to understand, devise and scale-up best learning and teaching practices to as many students as possible. This ambition has been bolstered most notably by the Bloom (1984) studies, which are still routinely cited throughout the AIED literature as a key justification and motivation for the field. This ambition had bootstrapped much of the work within the field and it has spurred in-depth research examining how specific populations of students learn, what are the prerequisites (cognitive, affective, and pedagogic) for successful learning, and how AIED technologies might be designed to help develop and capitalise on such learning prerequisites.

Personalisation through adaptivity of assessment and feedback (for the purpose of this article used in the broad sense of pedagogical support) remains at the heart of the work conducted by AIED researchers, regardless of their specific areas of specialisation, or their philosophical or epistemological perspectives. This is why, to date, the AIED community repeatedly voted to retain its long-debated connection with the wider field of AI – a domain like AIED insofar as its central paradigm of adaptive agent technologies, but unlike AIED as far as its aim to emulate human capacities *only* to the extent that it is useful to a given application's success in achieving its specific goals. By contrast to many other subfields of AI, the fidelity of AIED applications to human cognition and behaviours remains highly relevant to the ability of its technologies to support human learning and development in cognitively congruent, efficacious and socio-culturally desirable ways (see also Rismanchian and Doroudi (2023) for an overview of the evolution of the role of AI in Education).

As such, in its ambitions, focus and outcomes to date, the AIED research arguably represents a unique sub-discipline of AI, wherein the technology is not so much intended to compensate for the lack of certain abilities in their users (e.g., radiographers' ability to spot accurately signs of cancer in x-ray images), or to bypass some abilities (e.g., navigating a driver through an unfamiliar geographical terrain), but rather it is to emphasise, help train and strengthen the capacities with which we are biologically endowed. Thus, in line with the overarching goals of education, AIED technologies are purposefully created as *interventions* to enhance and change *human* thinking and behaviour. In this, the AIED field – its research, specific technologies, ambitions and funding – reflects certain policies and ideologies behind the different intervention designs (Madaio et al., 2022). And thus, the AIED research and practices are inevitably intertwined with the wider social, cultural, economic and political contexts of the technological applications created by the field. Any evaluation of the field's track record, of current trends and achievements, and of future directions, also requires awareness of and involvement with the broader landscape of societal needs and related ethical challenges.

Following almost half a century-long history, the AIED Community has been busy addressing criticisms of being narrowly focused on a drill and practice pedagogical paradigm, by pro-actively exploring, developing and evaluating a diversity of approaches to learning and teaching with AIED technologies. Some examples of this continually changing and expanding focus include research and systems utilising exploratory learning, learning by teaching, collaborative learning and self-regulated learning, as well as a growing focus on a multitude of subjects from STEM to art, to human social interaction and communication. Furthermore, in contrast to the long dominating view in traditional formal education that learning is predominantly a cognitive endeavour, the field is also an early pioneer of research on factors such as context, culture, personality, gender, affect and motivation as fundamental to learning, with important contributions to the development of relevant theories and of AIED designs having been made at least since the late 1980s. In this, the field affords a unique human-centred perspective for the wider AI research, by considering questions about user experience as integral to human learning, and as intricately dependent on technology-external contextual factors. Among other things, this perspective is reflected in the AIED community having been an early adopter of methodologies that integrate participatory, user-centred design and action research with engineering, to ensure the relevance of the experiences and support offered by different AIED applications.

Importantly, AIED is also a sub-field of AI which recognised and addressed early on the need for transparency, accountability and flexibility of AI designs as a means to mitigate against the inevitable lack of absolute accuracy in the technology's detection and interpretation of human behaviours and internal psychological states (Bull & Pain, 1995; Beck et al., 1997). Notably, this recognition has resulted in research and development of Open Learner Models (OLMs) and systems that *invite the users in* (Bull & Kay, 2006), whereby the users are given different degrees of agency, control and ownership over the data generated and processed by AIED systems, and over their learning processes and experiences. With this breadth of interdisciplinary perspectives providing the basis for much of AIED's endeavours, the field's coming of age could be said to be manifest in the AIED Community's ongoing recognition of the necessity to explore diverse ways in which AIED technologies might guide and support how different learners learn, and in its increasing acknowledgement of the importance to design the technologies and uses thereof with the entire learning and teaching ecosystem in mind.

Despite these cumulative achievements many gaps and blind spots remain. At least two such gaps relate to:

1. the Community's marked disengagement from the debates and actions around responsible AI, and its reluctance to define its own role in informing and steering those debates, especially as relates to guiding the EdTech industry and policies;
2. its persistent favouring of technologies targeting dominant groups of learners – a persistence that reinforces the field's historic perspective on education as a top-down and systemically sanctioned process of remediation of deficits in learners' knowledge and competencies.

I believe that recognising and eventually addressing these blind-spots as part of the field's explicit agenda is necessary to allow the Community to elevate its research beyond what many outsiders consider a peripheral cottage industry. In turn, this will likely open to the Community the possibility of a greater and more directed influence over the wider AI agenda, the best practices at the front line of education, as well as AIED and EdTech industry and policies.

## AIED Within the Broader Responsible AI Agenda

Although AIED's ambitions have been always and explicitly connected with front- line education, the field's engagement with broader implications of AI for human functioning, learning and socio-cultural impact remains relatively limited. There is an emergent acknowledgement within the AIED Community of the wider agenda on responsible AI, notably marked by a series of recent publications on the Ethics of AI in Education (Holmes & Porayska-Pomsta, 2022; Baker & Hawn, 2021) that attempt to contextualise and operationalise the ethics of AI for educational applications. However, much of the dialogue about the ethical issues in AIED and its role in contributing to the wider debates remains at the periphery of the Community's main interests (e.g., see the AIED conference 2022 panel discussion on AIED's Coming of Age; Holmes et al., 2021). In this sense, the responsibility for setting the agenda for responsible AIED seems to be delegated to other actors in the EdTech and the policy of education sectors. This blind spot has been flagged and discussed at some length in Porayska-Pomsta et al. (2023), where the Community's complacency regarding its responsibility to help steer in a coordinated way the EdTech and related education and market policies has been highlighted alongside some likely causes.

Regardless of what these causes actually are, the Community's so far limited participation in the wider responsible AI efforts reflects on its (im-)maturity with respect to its own understanding of the critical role that it must play in disseminating and actioning its collective knowledge. Such an understanding is necessary to promote the relevance of AIED's contributions to the wider AI developments, educational research and practice, AI and EdTech industry, and to the corresponding governance decisions. It is also necessary to safeguard against potential malpractice and monetisation thereof by agents who may pay limited attention to foundational questions about the purpose of education, or the nature of learning, or to evidence- based practices, but who would be happy to appropriate the AIED banner in the name of their own interests or politik. In this context, it seems both pointless and irresponsible to distance ourselves from the EdTech industry as a separate and variably informed sector, or to surrender our influence over the policies that endorse this industry.

A curious manifestation of the Community's disengagement with the issues at hand is the frustration voiced by some of its members during the AIED 2022 conference regarding the role and the use of data in the context of AIED research. Specifically, it seems that the questions about the risks associated with a possibility of education-focused data exploitation for life-long surveillance of individuals within broader social and political systems are sometimes dismissed as over-zealous ethics mongering and as ignorant about the true and noble aims of the AIED field, which is to drive AIED's systems' decision-making and adaptivity and help make the world a better place. After all, accessing, generating and analysing data about learners is necessary to our ability to tailor learning supports to individual learners. Here, an overt assumption has been articulated that as long as these aims are made clear and as long as we adhere to the institutional research ethics requirements and secure consent from the research participants, then we can consider ourselves to be doing both ethical things and to be doing them ethically.

Yet many examples from wider AI show that good intentions conceived in research labs often do not extend to the mainstream uses. Apart from the now well- documented issues with diverse forms of bias encoded in data and algorithms, there are also examples of applications being used differently than intended by their designers. Examples of such discrepancies in educational contexts range from the way that data and algorithms have been appropriated and misused with disastrous consequences at the systemic policy levels (e.g., O'Neill, 2016), to more AIED specific scenarios of decision-making applications potentially serving to confirm pre- existent biases, e.g., of teachers (Holstein & Doroudi, 2022). To consider and to understand the potential risks and misuses of the research and outcomes of a field will not lead to the diffusion of its aims and ambitions, or to the field 'losing its way'. Rather, it is likely to strengthen its relevance and contributions, while also offering new perspectives on what might be needed on the ground and on how such needs might be addressed in ways that can be valuable or even optimal, but otherwise obscured.

## Dominant Epistemologies and Methodologies

Related to the need for the community to expose its ambitions to new perspectives and thinking is the AIED's predominant focus on mainstream education and by extension – on dominant target learners, curricula and learning supports / pedagogies (the second gap identified above – see also Baker and Hawn (2021), Porayska-Pomsta et al. (2023), Treviranus (2022), Madaio et al. (2022)). In the early days of the field, such a focus was justified by the limitations presented by the fledgling computer technologies, a dominant model of education which was unchallenged by the Internet or social media, and by a general lack of recognition in the pre-woke culture of the differences between people being a sign of diversity to be embraced and to be inspired by, rather than a deficit to be ashamed of and to be eradicated. While there are some examples of AIED efforts towards neurodiverse populations, work focusing on physical disabilities, or learners from minority groups remains sparse within the field (Baker & Hawn, 2021; Porayska-Pomsta et al., 2023). Worryingly, while many fields have shifted towards treating different abilities as strengths and as starting points for interventions (e.g., Porayska-Pomsta & Rajendran, 2019), the predominant language used within AIED still casts such differences as *dis*-abilities (see e.g., the language used in the well intended Baker and Hawn (2021) paper), suggesting that the understanding and the mind-set towards *different-ableness* within the field is lagging behind other related disciplines of research and practice.

The *deficit model* of education adopted in many prominent examples of AIED systems, whereby learners are being *diagnosed* like medical patients for conceptual bugs that must be remediated or for missing competencies that must be instilled in them, remains the most common within AIED systems and the most prominent in the eyes of those outside the Community who care to look (Treviranus, 2022). Despite numerous improvements in how the different learning supports delivered by AIED systems might be tailored to give learners greater agency over their learning (e.g., through facilitating pro-active help-seeking by students, students' self-reporting on their emotional states, in collaborative or even exploratory settings), this deficit model falls short in its ability to promote inclusive practices, or to inspire methodological and technological advances in the field, or pedagogical innovation more broadly. In assessing learners through *overlaying* their knowledge and competencies onto those of experts or onto systemically prescribed curricula, this model relies heavily on patterns of typicality as well as on access to relevant, voluminous data: consider, for example, that in many autism-focused studies that are recognised and published by the experimental fields of cognitive psychology, a sample of 40, i.e. below a level needed for a statistically meaningful analysis, is considered large due to the idiosyncratic nature of this neuro-environmental condition and related needs. Not only patterns of typicality may not represent all students' situated strengths or needs, the focus on typicality is likely to lead to inflexible technological designs, whereby instead of technology adapting to the users, the users have to adapt to the technology – a phenomenon already researched by some (e.g., Shin et al., 2020), but whose transformational impact (for better or worse) on human thinking and behaviour remains an open question (see also Porayska-Pomsta & Holmes, 2022).

By contrast, exposure to both the needs and the contexts, in particular, of *differently abled* learners puts in sharp relief the limitations of the overlay modelling paradigm by challenging our assumptions about what actually constitutes typicality, our *a priori* ability to predict student behaviours accurately, and to get the necessary and sufficient data in the first place (see also Baker & Hawn, 2021). Lack of such exposure precludes our experience of (not so atypical) contexts and situations where learning and the corresponding pedagogical and transactional supports are less a matter of generalising theoretically good practices, instead being more a matter of highly informed pedagogical improvisation and of teaching that is reactive as well as sensitive to the often volatile learner states and learner readiness to engage (e.g. Prizant et al., 2006; Porayska-Pomsta, 2016).

The issues with accessing voluminous data in special educational needs contexts also have implications for the research methods that are often used therein. In the context of neurodiverse ableness, in particular, qualitative methods and interpretivist approaches are often a must for enabling access to lived experiences of individual learners and to behaviours that are outliers to the 'typical' patterns (Ruttenberg et al., 2023). It is in these outlier data that practitioners and researchers often find keys to optimal support regimes. Such approaches do not diminish the scientific rigour of the research conducted in this way. But they are often the necessary accessories to engineering knowledge in contexts that are high stakes for the individuals. The extent to which such methods alone can inform generalisable conclusions and standardised approaches remains an open question. However, what they do spotlight is the necessity to extend our field of view, to be inspired by a diversity of best possible practices, and to consider certain dimensions, such as user control over the operation of the systems we build, as critical to the adaptive potential of those systems (see also, Bull & Kay, 2016 for an AIED centric perspective on the questions of user control).

Despite all of this, the qualitative and interpretivist approaches are often met with scepticism by many members of the AIED Community, who favour quantifiable research outcomes. This tendency is reflected in research contributions typically accepted both to the annual AIED conference and to the Society's Journal, with reviewers evidently tending towards attributing greater value to data-driven, controlled experimental designs over other possible and at times more appropriate approaches (see again Rismanchian and Doroudi (2023) for an overview of the related AIED trends over the decades). While this scepticism is understandable, its consequence is that many researchers working on AI for non-mainstream education and who produce valuable AIED-relevant research are likely to seek other venues to publish their work. Furthermore, although the deficit model serves to some extent the existing mainstream practices in well-defined domains, it likely hampers the Community's ability to critique, challenge and innovate those practices by reference to the perspectives of learners who are disenfranchised by the mainstream education, or who are set up for failure in highly standardised educational settings. Theirs is a perspective that is increasingly recognised in other fields as critical to our being able to respond to the changing global education landscape and needs of learners. As such this perspective ought to be considered as foundational to the AIED field's ability to remain relevant and to being able to extend its ambition from the need to do the known things better (e.g., its obsession with Bloom's 2 sigma effect), to also considering how to do better things differently (e.g., helping steer and deliver education towards inclusion and diversity-oriented practices and policies). Importantly, adopting this wider and more inclusive focus might help in changing the perceptions of AIED held by many outsiders from being a field that is solely dedicated to K-12 education, to being acknowledged for its substantial work on life-long human learning and development, and for its fundamental research on human cognition.

**Toward a Manifesto for a Pro-Actively Responsible AIED**

The gaps spotlighted above stand in contrast both with original motivation for the field and with many individual examples of research produced by the Community of pedagogically innovative approaches (from Cognitive Tutors, to OLMs, to Teachable Agents, and more) that are both educationally feasible and efficacious. These examples ought to be of great interest to wider educational practices and policies. Unfortunately, few people outside the immediate Community know about either those examples or about AI in Education as a mature field of research. Increasingly, researchers and entrepreneurs from other disciplines and walks of life come up with seeming eureka ideas about AI's potential for education, often without the backing of the decades of thinking and experimentation that AIED research to date provides. Some slightly better-informed researchers in other AI subdisciplines have long ago dismissed AIED as being confined to the pre-history of GOFAI ('*good old-fashioned AI*') era of expert systems.

Yet, many of the examples of ideas, system designs and evidence-based practices can be used straight off the proverbial AIED research shelf to challenge, to inform, and to steer broader AI developments and agenda at technical as well as policy levels, and with tangible benefits for AIED specific ambitions such as dedicated funding. For example, in the context of recent broader AI discussions around the lack of transparency, accountability, and human-centredness of AI, the most obvious examples of innovations delivered by AIED are those which represent a potential challenge to the *standard model of AI*, i.e., a model where AI has a full and difficult to track or to challenge control of the interaction with the human. In this standard model AI is a fully autonomous agent whose job is to single-mindedly find ways that maximise the achievement of a pre-set objective, irrespective of the users' real-time or long-term needs or contexts.

In his popular science book entitled "Human Compatible", Russell (2019) highlighted the issues with this standard model of AI as residing in its rigid insistence on solving problems within local optima of the limited contexts within which AI operates, and in human fallibility in specifying the kinds of objectives that can be guaranteed to be ethical and beneficial to all possible stakeholders at all times. In this book, and in subsequent high-profile talks, he proposed a brilliant idea for changing the standard model of AI to a more flexible one that is sensitive to human needs and contextualised preferences – a model that is open to manipulation and personalisation by their users, and a model that gives data back to the users and whose objectives can be negotiated.

To anyone familiar with Open Learner Models and with their origins lying in the need to improve the accuracy of student modelling, this likely sounds both encouraging and frustrating. It is encouraging, because it demonstrates that hard-core AI researchers and thinkers like Russell are beginning to think about how to do better things with different models of AI to ensure human safety and wellbeing. They are starting to think about human agency in the human-AI interaction as underpinning AI for human good. It is frustrating, because the AIED Community have not only come up with similar proposals, but also have been developing and evaluating their different instantiations for efficacy and value to users since the early 1990s. Yet, in the context of broader AI, it is as if this work does not exist. Despite the approaches devised, trialled and evaluated to date by the AIED research Community having always originated with the questions about AI's specific benefits to the human users (sic. the learners) and despite them representing pre-eminent examples of relevance to informing new forms of AI design and use, they remain mostly unknown to the wider AI work. As a community we do need to ask ourselves why this is so, and what, if anything, do we want to do about this.

As someone who cares deeply about AIED as a field and finds her professional home among its many talented experts from around the World, I believe that for the Community's own survival, there is a pressing need for our coordinated engagement with broader issues outside our immediate and traditional research foci and comfort zones. This is particularly pressing given the emergent landscape of the *economics of AI* and the increasing global trends towards commercialisation of public good institutions (see e.g., Mazzucato et al., 2022; Mazzucato, 2018, 2021). These trends stand in direct opposition to the original motivation and ambitions for AIED. The policy makers concerned with the way that AI will impact society at different levels are looking for evidence-based guidance on what AI best practices might look like, as well as to safeguard against risks and harms of AI in different contexts including Education. The broader AI research and industry is actively involved in the questions about how to do better things with AI better, in order to deliver long term benefits for society at a collective and individual levels.

Questions surrounding AI in Education are both high-stakes and central to the welfare of current and future generations. Yet the work of the AIED Community is still at best a footnote for broader AI, related policies and governance of EdTech. It is our responsibility as a scientific community to pro-actively seek any opportunity to inform and help steer policy decisions and best practices based on the collective and cumulative knowledge generated through our research over the past five decades. Otherwise, the fact that we as a Community place ourselves on a high pedestal of greater knowledge and morally laudable ambitions will not shield us from being complicit in our disciplinary banner being used for low quality products or bad policy decisions related seemingly to AI in Education. By not engaging directly and decisively with the issues at hand, we are agreeing to AIED's research accomplishments and findings being largely irrelevant to the decisions being made by other, and likely less informed, sectors.

The AIED's coming of age has to be about considered self-awareness of our own accomplishments and of our critical relevance to agenda setting for human-centred AI and educational innovation. The following 5-point manifesto provides a start towards the action needed:

1. **Inform and Challenge**: AIED Community must play a pro-active role in disseminating and actioning its collective knowledge beyond its own Community's comfort zone to enhance its relevance to wider AI developments, educational research and practice, AI and EdTech industry, and corresponding governance decisions. AIED Community can and should challenge, inform, and steer broader AI developments and agenda at technical as well as policy levels, in order to achieve its long-standing aspirations of delivering best education to all, while also offering safeguards against educationally shallow and potentially harmful applications;
2. **Expand the field of view**: The AIED Community needs to expand its understanding of the broader implications of AI for human functioning, learning, and socio-cultural impact to identify the potential risks and misuses of its research, and outcomes and to strengthen its relevance and contributions beyond its own boundaries. Questions about data exploitation for life-long surveillance of individuals within broader social and political systems need to be raised along with questions about the transformative impact (for better or for worse ) of AI on innate human capacities that were historically considered critical;
3. **Engage with, guide and hold to account** the wider EdTech industry and the policies that endorse it: AIED should not distance itself from the EdTech industry as a separate sector or surrender its influence over the policies that endorse it. It should encourage and work towards providing auditing mechanisms and examples of best practice in the field against which the quality of all AIED endeavours, products and uses thereof might be evaluated;
4. **Examine the good intentions for possible implications** beyond its intended applications: The AIED Community should understand that good intentions conceived in research labs often do not extend to mainstream uses and that data and algorithms can encode diverse forms of bias and risks to individuals and groups. Understanding the potential risks and misuses of AIED research will not lead to the diffusion of its aims or its ambitions, but will likely strengthen its relevance and contributions and will help innovate AIED practices and technologies;
5. **Put your money where your mouth is**: The AIED Community needs to expand its focus on mainstream education and dominant target learners, curricula, and learning supports/pedagogies to explore its potential for learners who are differently abled or who may be otherwise disenfranchised by the mainstream educational systems and curricula. A shift towards treating different abilities as strengths and embracing diversity in learners will advance the field towards its ambition to contribute to diversity and inclusion and to deliver good education to all. It will also open new possibilities for AIED and broader AI research and technological innovation, by offering a challenge and alternatives to the standard model of AI whose goal is to optimise the environment (including the human mind) through habituating the users to particular and often limited forms of perception, thinking and (inter-)action.